\documentclass[acus]{JAC2000_mc} 
\usepackage{graphicx}
\begin{document}

\thispagestyle{empty}

\onecolumn

\begin{flushright}
{\large
SLAC--PUB--8871\\
June 2001\\}
\end{flushright}

\vspace{.8cm}

\begin{center}

{\LARGE\bf
Study of Beam Energy Spectrum Measurement\\
in the NLC Extraction Line~\footnote
{\normalsize
{Work supported by Department of Energy contract  DE--AC03--76SF00515.}}}

\vspace{1cm}

\large{
Y.~Nosochkov and T.O.~Raubenheimer\\
Stanford Linear Accelerator Center, Stanford University,
Stanford, CA 94309}

\end{center}

\vfill

\begin{center}
{\LARGE\bf
Abstract }
\end{center}

\begin{quote}
\large{
The NLC extraction line provides a secondary focal point with a low $\beta$
function and 2~cm dispersion which can be used for measurement of the
beam energy spectrum.  In this study, tracking simulations were
performed to transport the 0.5~TeV electron beam from the
Interaction Point (IP) to the secondary focus (SF), ``measure'' the
resultant transverse beam profile and reconstruct the disrupted IP energy
spread.  In the simulation, the obtained energy spectrum reproduced the
initial IP spread reasonably well, especially with the vertical dispersion
at SF which provides larger ratio of dispersion to the betatron beam
size. More details of this study can be found in Ref.~\cite{long}.
}
\end{quote}

\vfill

\begin{center}
\large{
{\it Presented at the 2001 Particle Accelerator Conference
(PAC 2001)\\
Chicago, Illinois, June 18--22, 2001}
} \\
\end{center}

\newpage

\pagenumbering{arabic}
\pagestyle{plain}

\twocolumn

\title
{STUDY OF BEAM ENERGY SPECTRUM MEASUREMENT\\
IN THE NLC EXTRACTION LINE~\thanks
{Work supported by Department of Energy contract 
DE--AC03--76SF00515.}\vspace{-6mm} }

\author{Y.~Nosochkov and T.O.~Raubenheimer\\
SLAC, Stanford University, Stanford, CA 94309, USA}

\maketitle

\begin{abstract}

The NLC extraction line provides a secondary focal point with a low $\beta$
function and 2~cm dispersion which can be used for measurement of the
beam energy spectrum.  In this study, tracking simulations were
performed to transport the 0.5~TeV electron beam from the
Interaction Point (IP) to the secondary focus (SF), ``measure'' the
resultant transverse beam profile and reconstruct the disrupted IP energy
spread.  In the simulation, the obtained energy spectrum reproduced the
initial IP spread reasonably well, especially with the vertical dispersion
at SF which provides larger ratio of dispersion to the betatron beam
size. More details of this study can be found in Ref.~\cite{long}.

\end{abstract}

\vspace{-1mm}
\section{INTRODUCTION}

In a linear collider, the strong beam-beam interaction generates
significant beamstrahlung.  For the high-energy physics experiment to make
optimal use of the luminosity, it is important to know the luminosity
spectrum.  This is done in the NLC design by measuring the energy spectrum
of the disrupted beam in the beam extraction line which transports the beam
from the IP to the beam dump.

\begin{figure}[tb]
\centering
\includegraphics*[width=55mm, angle=-90]{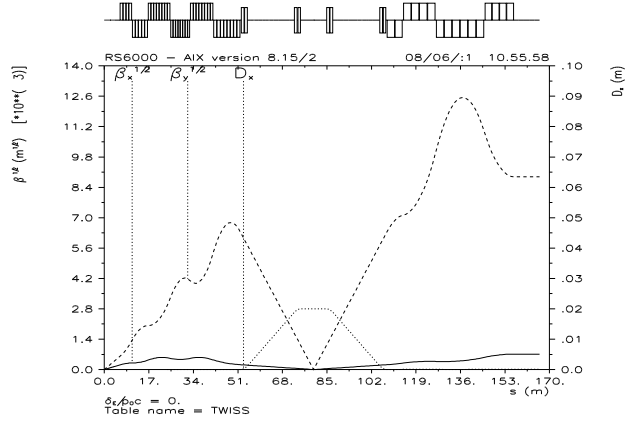}
\vspace{-1mm}
\caption{Extraction line lattice functions.}
\label{beta}
\vspace{-3mm}
\end{figure}

The present design of the NLC extraction line optics ~\cite{dump1,dump2} is
shown in Fig.~\ref{beta}, where the beam travels from the IP (on the left)
to the dump.  The optics contains two multi-quadrupole systems, where the
first system performs a point-to-point focusing from the IP to a secondary
focus (SF), and the second system generates a parallel beam at the dump.
Between the two quadrupole sets there is a symmetric four bend chicane
generating 2~cm displacement and dispersion at the SF.

To accurately measure the energy spectrum of the disrupted beam, the
dispersive beam size $\eta\delta$ should be large compared to the betatron
beam size $\sqrt{\beta(\delta)\epsilon}$.  The original optics was designed
with the horizontal chicane as shown in Fig.~\ref{beta}, but vertical bends
may be used as suggested by K.~Kubo of KEK~\cite{kubo} to improve the
resolution.

The on-energy $\beta$ functions at the SF can be derived from the IP
$\beta^*$ values and linear matrix $R_{ij}$ between IP and SF:

\vspace{-6mm}
\begin{eqnarray}
R_{12}\!=\!0,\ \ 
R_{11}=-4.5233,\ \ 
\beta_{x}=R_{11}^{2}\beta_{x}^{*}, \nonumber \\
R_{34}\!=\!0,\ \ 
R_{33}=-0.4549,\ \ 
\beta_{y}=R_{33}^{2}\beta_{y}^{*}.
\label{eq-beta}
\end{eqnarray}
\vspace{-6mm}

In this study, we used one particular set of the NLC beam
parameters~\cite{param} listed in Table~1.  These parameters correspond to
the disrupted 0.5~TeV beam at the IP.  The beam disruption occurs in the
collision and significantly increases the beam divergence, emittance and
energy spread.  The disrupted distribution at the IP was obtained using
GUINEA--PIG code~\cite{gpig,kathy}, and the corresponding emittance and
lattice functions were reconstructed from this distribution as shown in
Table~1.

According to Eq.~\ref{eq-beta} and Table~1, for a fixed 2~cm dispersion,
optics with the vertical chicane provides a larger ratio of the dispersion
to the same plane betatron size at the SF, and therefore should result in
better accuracy in the energy spectrum measurement.  Note, that
$\beta(\delta)$ at SF can grow significantly with $\delta$ due to the shift
of $\beta$ waist.

Below we compare tracking and measurement simulations for the extraction
line with the horizontal and vertical 2~cm chicane.  The particle tracking
was done using a version of DIMAD code with accurate handling of large
energy errors~\cite{dimad}.  Effect of the corrected 6~T detector solenoid
is included.

\begin{table}[hb]
\vspace{-4mm}
\begin{center}
\caption{Disrupted beam parameters at IP.}
\medskip
\begin{tabular}{ll}
\hline
Emittance, $\epsilon_x/\epsilon_y$ (m$\cdot$rad) [$10^{-13}$] 
                                        & 120 / 1.02 \\
Beam size, $\sigma^*_x/\sigma^*_y$ (nm) & 198 / 3.2 \\
Divergence, $\sigma^{*\prime}_x/\sigma^{*\prime}_y$ ($\mu$rad) 
                                        & 125 / 33 \\
$\beta^*_x/\beta^*_y$ (mm)              & 3.259 / 0.103 \\
$\alpha^*_x/\alpha^*_y$                 & 1.805 / 0.306 \\
Energy per beam (GeV)                   & 523 \\
Particles per bunch                     & $0.75 \times 10^{10}$ \\
Bunches per train                       & 95 \\
Repetition rate (Hz)                    & 120 \\
Disruption parameter, $x/y$             & 0.094 / 6.9 \\
Average energy loss per particle        & 9.5\% \\
\hline
\end{tabular}
\end{center}
\vspace{-5mm}
\end{table}

\vspace{-1mm}
\section{SIMULATIONS}

\begin{figure}[tb]
\centering
\includegraphics*[width=75mm]{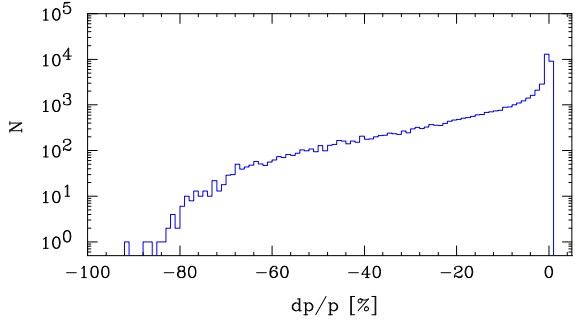}
\vspace{-3mm}
\caption{Disrupted beam energy spread at IP.}
\label{dndp-ip}
\vspace{-3mm}
\end{figure}

The GUINEA--PIG code was used to generate $5\!\cdot\!10^4$ macro-particles
to represent the disrupted distribution at the IP.  As shown in
Fig.~\ref{dndp-ip}, this distribution has a huge energy spread.  The
disrupted beam then was tracked to the secondary focus using DIMAD.

The small betatron beam size and non-zero dispersion at the SF result in a
significant correlation between a particle transverse position $x$ (or $y$
for vertical $\eta$) and energy.  This makes it possible to reconstruct the
beam energy spectrum based on the beam profile measurement at the secondary
focus.  Neglecting synchrotron radiation, a particle coming into chicane on
axis will have a transverse deflection

\vspace{-6mm}
\begin{eqnarray}
x_{\eta}=\frac{\eta\delta}{1+\delta}
\label{eq-x}
\end{eqnarray}
\vspace{-5mm}

\noindent at the secondary focus.  Equation~\ref{eq-x} can be used to estimate
the particle energy deviation $\delta$ based on measured $x$ (or $y$) and
known $\eta$ at the secondary focus:

\vspace{-6mm}
\begin{eqnarray}
\delta=\frac{x}{\eta-x}.
\label{eq-delta}
\end{eqnarray}
\vspace{-5mm}

Using Eq.~\ref{eq-delta}, one can also convert measured beam profile $N(x)$
into the energy spectrum $N(\delta)$.  Equation~\ref{eq-delta} is only correct
if $\delta$ is constant and the deflections are caused entirely by
dispersion.  In reality, several other factors contribute to particle
position at the secondary focus:
\vspace{-0mm}
\begin{itemize}
\vspace{-2mm}
\item
Betatron motion $\sim\sqrt{\beta(\delta)\epsilon}$.
\vspace{-2mm}
\item
Synchrotron radiation causing  random energy loss. 
\vspace{-2mm}
\item
Quadrupole misalignment and bending field errors.
\vspace{-2mm}
\item
Incoming beam offsets $x^*$, $y^*$ at the IP.
\vspace{-2mm}
\end{itemize}
\vspace{-0mm}
Measurement errors have to be taken into account as well.
Denoting the above contributions as $\Delta x$, particle deflection at 
the SF can be expressed as $x\!=\!x_{\eta}\!+\!\Delta x$.
Clearly, an accurate estimate of $\delta$ in Eq.~\ref{eq-delta} requires that
$|\Delta x|\!\ll\!|x_\eta|$.

\vspace{-1mm}
\subsection{Energy Resolution Analysis}

\begin{figure}[tb]
\centering
\includegraphics*[width=80mm]{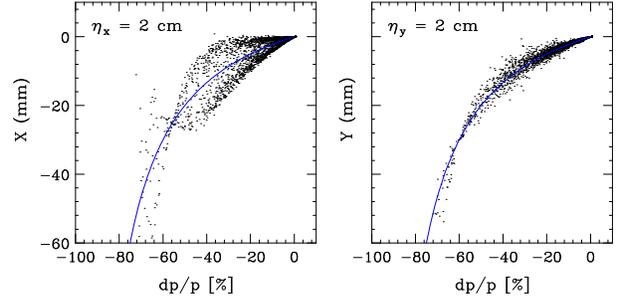}
\vspace{-3mm}
\caption{$x$ and $y$ distributions vs. $\delta$ at the SF
for horizontal and vertical chicane, respectively.}
\label{p1-xy}
\vspace{-3mm}
\end{figure}

To verify the actual dependence of particle positions on energy, the
disrupted beam of $5\!\cdot\!10^4$ particles was tracked from IP to SF.
The simulation included synchrotron radiation effects, but no magnet errors
were used.  The resultant $x$ and $y$ distributions versus $\delta$ at the
SF are shown in Fig.~\ref{p1-xy} for 2~cm horizontal and vertical chicane,
respectively.  The solid line in Fig.~\ref{p1-xy} is the analytic
displacement in Eq.~\ref{eq-x}, and $\frac{\Delta p}{p}$ is the initial
energy error at the IP.  Note that synchrotron radiation between IP and SF
reduces average particle $\delta$ at the SF by $\sim\!1.5\!\times\!10^{-3}$
compared to IP $\delta$.

As shown in Fig.~\ref{p1-xy}, the particle $\Delta x$ spread versus
$\delta$ is wider in optics with the horizontal chicane due to a larger
$\sigma_x$.  Therefore, the reconstruction of beam energy spectrum based on
beam profile measurement should be more accurate with the vertical chicane
where the dispersive contribution (Eq.~\ref{eq-x}) is dominant.

\begin{figure}[b]
\vspace{-1mm}
\centering
\includegraphics*[width=60mm]{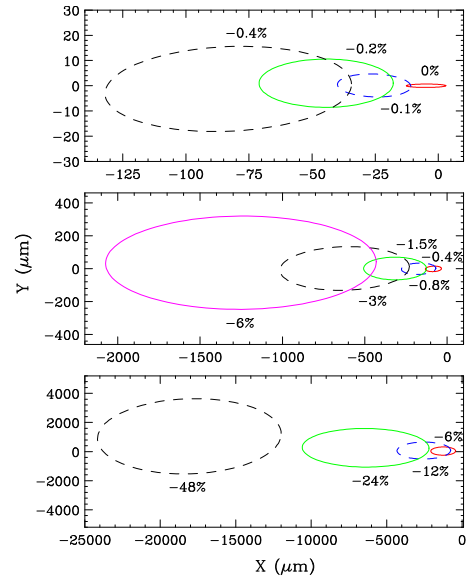}
\vspace{-3mm}
\caption{Monoenergetic ellipses at the SF with $\eta_x\!=\!2$~cm.}
\label{ellpsx}
\vspace{-1mm}
\end{figure}

\begin{figure*}[tb]
\centering
\includegraphics*[width=140mm]{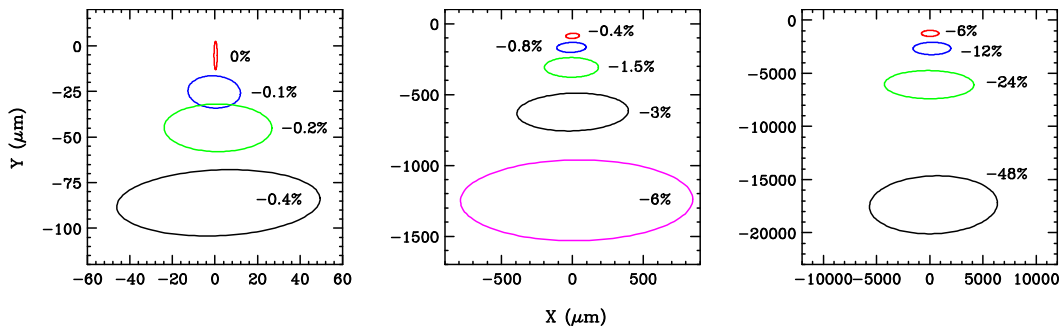}
\vspace{-3mm}
\caption{Monoenergetic ellipses at the secondary focus with 
$\eta_y\!=\!2$~cm.}
\label{ellpsy}
\vspace{-3mm}
\end{figure*}

The energy resolution can be also examined using analysis suggested by
Kubo~\cite{kubo}.  In this method, the simulated beam at SF is divided
into almost monoenergetic slices with different energy which are then
evaluated in terms of $x$ and $y$ size, position and orientation.
Fig.~\ref{ellpsx} and~\ref{ellpsy} show these slices at different $\delta$
in the form of one sigma ellipses on the $x$-$y$ plane.  Clearly, the
resolution between the slices is better in the vertical chicane optics.

Note that the ellipse $\delta$ in Fig.~\ref{ellpsx} and~\ref{ellpsy} is the
initial IP $\delta$.  Synchrotron radiation randomly reduces the energy
which distorts particle deflections at the SF, but relatively it affects
more the particles with $\delta\!\sim\!0$.  This results in a
disproportionally wide ellipse in the direction of dispersion at
$\delta\!=\!0$.  This effect is negligible for $|\delta|>1\%$.

Further analysis~\cite{long} shows that the ellipse size at SF is
approximately proportional to $|\delta|$, which can be interpreted as
$\beta(\delta)\!\sim\!\delta^2$, and that this dependence mostly comes from
the shift of $\beta(\delta)$ waist.  Analysis of the ellipse size and
position shows that the ratio of an average particle displacement at SF to
the beam size as a function of $\delta$ is almost a factor of 3 larger with
the vertical dispersion, and therefore is better for energy resolution.

\vspace{-1mm}
\subsection{Energy Spectrum Measurement}

In the simulation, the beam profile measurement at the secondary focus was
done using 50~$\mu$m steps over the range of 25~mm in the direction of
dispersion to simulate a wire scanner.  The particles were collected for
each of the 50~$\mu$m bins ($x$ or $y$), and the beam profile histogram
$N(x)$ was generated, where $N$ is the number of particles per bin.  The
range beyond 25~mm was not considered due to low statistics, and no
measurement errors were included.

The beam energy spectrum $N(\delta)$ can be obtained from the $N(x)$
profile by converting the $x$ (or $y$) bins into $\delta$ bins.  According
to Eq.~\ref{eq-delta}, the width of $\delta$-bin varies with $x$ as

\begin{figure}[htb]
\centering
\includegraphics*[width=75mm]{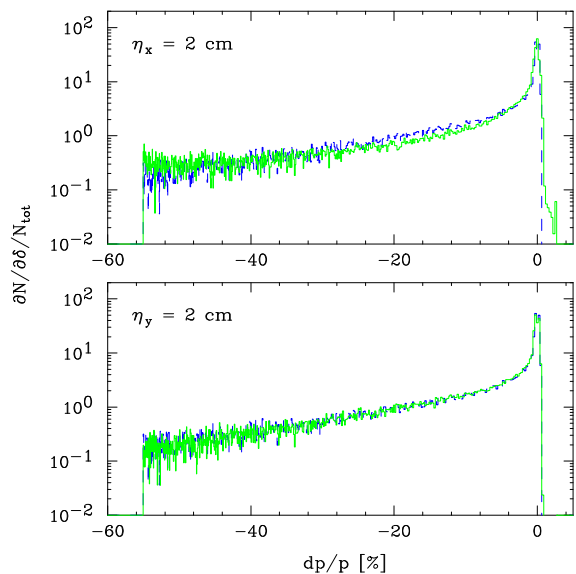}
\vspace{-3mm}
\caption{Original (blue, darker shade) and measured (green) energy
spectrum $\frac{1}{N_{tot}}\frac{dN}{d\delta}$ for $x$ and $y$ chicane.}
\label{dndp-l}
\vspace{-3mm}
\end{figure}

\vspace{-5mm}
\begin{eqnarray}
\Delta_{\delta}=\frac{\eta}{(\eta-x)^2}\Delta_{x},
\label{eq-deltabin}
\end{eqnarray}
\vspace{-4mm}

\noindent where $\Delta_x\!=\!50$~$\mu$m is the $x$ (or $y$) bin width.  At
2~cm dispersion, the $\Delta_{\delta}$ width gradually reduces from 0.25\%
at $x\!=\!0$ to 0.05\% at -25~mm.  To avoid dependence on the bin width, we
normalized $N(\delta)$ to the corresponding width $\Delta_{\delta}$.  For a
more general result, we also normalized $N(\delta)$ to the total number of
particles $N_{tot}$ in the histogram.  The resultant energy distribution
$\frac{1}{N_{tot}}\frac{dN}{d\delta}$ was compared with the initial
spectrum at the IP.  Both histograms are shown in Fig.~\ref{dndp-l} for the
horizontal and vertical chicane, where the blue line (darker shade) is for
the original IP spectrum and the green line for the ``measured'' spectrum
at SF.  The $\delta$ range in Fig.~\ref{dndp-l} is limited at about
$\delta\!=\!-55\%$ due to 25~mm range used in the beam profile measurement.

Comparison of the histograms in Fig.~\ref{dndp-l} shows that optics with
the vertical chicane provides a more accurate reconstruction of the initial
IP energy spectrum.  A closer view shows that the vertical measurement even
reproduces the incoming double peak profile near $\delta\!=\!0$, while the
horizontal histogram is not accurate in this range.  In the horizontal
spectrum, there are some particles in $\delta>0$ range that are not
present in the initial IP distribution.  This is the result of larger
horizontal betatron oscillations which are interpreted as positive $\delta$
in Eq.~\ref{eq-delta}.  This effect, though, is somewhat exaggerated in
Fig.~\ref{dndp-l} due to logarithmic scale.

The most uncertainties in the measured energy spectrum appear near
$\delta\!=\!0$ and at the very low energy tail.  At small $\delta$, the
Eq.~\ref{eq-delta} may be not accurate due to relatively large betatron
oscillations and effects of synchrotron radiation energy loss.  At the very
low energies, accuracy is reduced due to low statistics in the beam tail
and smaller $\Delta_{\delta}$ width.  According to Eq.~\ref{eq-deltabin},
one could obtain constant $\Delta_{\delta}$ distribution for a beam profile
measurement with $\Delta_{x}\!=\!\frac{(\eta-x)^2}{\eta}\Delta_{\delta}$.

\vspace{-1mm}
\section{CONCLUSION}

Tracking simulations and beam analysis in the NLC extraction line show that
a beam profile measurement at the secondary focus can be used to
reconstruct the disrupted beam energy spectrum.  Optics with the vertical
chicane is preferred because of the smaller ratio of betatron size to
dispersion and therefore better energy resolution.

\end{document}